\documentclass[
  aps,
  prd,
  reprint,
  amsmath,amssymb,
  superscriptaddress,
  nofootinbib,
  longbibliography
]{revtex4-2}

\usepackage{graphicx}
\usepackage{bm}
\usepackage{xcolor}

\usepackage[textsize=tiny,colorinlistoftodos]{todonotes}

\newcommand{\dd}{\mathrm{d}}
\newcommand{\dv}[2]{\frac{\mathrm{d}#1}{\mathrm{d}#2}}      
\def\qty(#1){\left(#1\right)}        

\usepackage{amsmath,amssymb,bm}
\usepackage{graphicx}
\usepackage{booktabs}
\usepackage[colorlinks=true,linkcolor=blue,citecolor=magenta,urlcolor=blue]{hyperref}
\AtBeginDocument{\let\hbar\hslash}

\def\doi#1{\href{https://doi.org/#1}{\color{blue}#1}}
\usepackage{tikz}
\usetikzlibrary{shapes.geometric,shapes.symbols}
\DeclareRobustCommand{\orcidicon}{%
	\begin{tikzpicture}
		\definecolor{orcidgreen}{HTML}{166B3A}
		\draw[orcidgreen, fill=orcidgreen] (0,0)
		circle [radius=0.16]
		node[white] {{\fontfamily{qag}\selectfont \tiny ID}};
	\end{tikzpicture}
	\hspace{-2mm}
}
\foreach \x in {A, ..., Z}{%
	\expandafter\xdef\csname orcid\x\endcsname{\noexpand\href{https://orcid.org/\csname orcidauthor\x\endcsname}{\noexpand\orcidicon}}
}

\renewcommand{\orcidicon}{%
	\begin{tikzpicture}
		\definecolor{orcidgreen}{HTML}{166B3A}
		\draw[orcidgreen, fill=orcidgreen] (0,0) circle [radius=0.16] node[white] {{\fontfamily{qag}\selectfont \tiny ID}};
	\end{tikzpicture}
	\hspace{-2mm}
}

\newcommand{\Msun}{M_{\odot}}
\newcommand{\Rb}{R_{\mathrm{B}}}
\newcommand{\Rd}{R_{\chi}}
\newcommand{\mB}{m_{\mathrm{B}}}
\newcommand{\md}{m_{\chi}}

\newcommand{\rhod}{\rho_{\chi}}
\newcommand{\pB}{p_{\mathrm{B}}}
\newcommand{\pd}{p_{\chi}}
\newcommand{\Mtot}{M_{\mathrm{tot}}}
\newcommand{\MB}{M_{\mathrm{B}}}
\newcommand{\Mcyl}{M_{\mathrm{cyl}}}
\newcommand{\GBT}{Gauss--Bonnet theorem}

\begin{document}

\title{Weak gravitational lensing by a dark-matter-admixed neutron star: a self-consistent two-fluid halo and the Gauss--Bonnet deflection angle}

\author{Yashmitha Kumaran\,\orcidY{}}
\email{yashmitha@astro.unam.mx}
\affiliation{Instituto de Astronom\'ia, Universidad Nacional Aut\'onoma de M\'exico, \\ AP 70-264, Ciudad de M\'exico 04510, M\'exico.}

\author{Il\'idio Lopes\,\orcidA{}}
\email{ilidio.lopes@tecnico.ulisboa.pt}
\affiliation{\href{https://ror.org/020hp3377}{Centro de Astrof\'isica e
		Gravita\c{c}\~ao (CENTRA)}, Departamento de F\'isica, \\
	Instituto Superior T\'ecnico (IST),
	\href{https://ror.org/01c27hj86}{Universidade de Lisboa (UL)}, \\
	Av.~Rovisco Pais~1, 1049-001 Lisboa, Portugal}

\date{\today}

\begin{abstract}
We compute the weak gravitational deflection of light by a neutron star that carries a self-consistent dark-matter component, in the regime in which the dark matter forms an \emph{extended halo} reaching beyond the baryonic surface. Two observations motivate this configuration. First, a dark-matter core confined within the baryonic radius leaves no distinctive lensing signature, since by Birkhoff's theorem the exterior is Schwarzschild with the total mass, so a ray passing outside the star feels only that mass and the compactness it implies. Second, the signature we seek resides in the complementary case, in which the ray genuinely traverses the halo, so that the surrounding density enters the optical geometry directly. We model the star by integrating the coupled two-fluid Tolman--Oppenheimer--Volkoff equations, with the baryonic and dark components interacting solely through gravity, and we obtain the deflection angle from the resulting external profile through the Gibbons--Werner construction of the \GBT. The defining feature of the approach is that the deflection is tied to a \emph{self-consistent} two-fluid profile rather than to a medium inserted by hand. We show that, for impact parameters smaller than the halo radius, the deflection departs measurably from the point-mass prediction, the deficit being governed by the dark-matter fraction and the halo extent, and we argue that the \emph{shape} of this deficit furnishes a geometric, lensing-based handle on the degeneracy between dark matter and the nuclear equation of state. The framework thereby unites the structural modelling of compact stars with their gravitational-lensing phenomenology.
\end{abstract}

\maketitle

\section{Introduction}
\label{sec:intro}

Neutron stars are the densest material objects known, and their interiors probe a regime of cold, charge-neutral, $\beta$-equilibrated matter that no terrestrial experiment can reach~\cite{LattimerPrakash2016,OertelEtAl2017}. Two largely separate lines of enquiry have grown up around them. The first concerns their structure and thermal evolution: the equation of state (EoS) of dense matter, the cooling driven by neutrino and photon emission, and the observables, namely masses, radii, tidal deformabilities, and oscillation frequencies, through which the EoS is constrained~\cite{Sagun2020,LattimerPrakash2016}. The second concerns their action as gravitational lenses: the bending of light in their strong external field, and what that bending records about the mass and the geometry through which the ray passes. This work unifies the two. We ask what the structure of a dark-matter-admixed neutron star imprints upon the weak deflection of light, and we compute that imprint self-consistently.

Interest in dark matter inside neutron stars is by now well established~\cite{Bertone2005,PanotopoulosLopes2017,BurasStubbsLopes2024}. Should a component of the cosmological dark matter be captured by, or co-assembled with, a compact star~\cite{GoldmanNussinov1989,PerezGarciaSilkStone2010,KouvarisTinyakov2011}, it modifies the star's macroscopic properties. The issue is that those modifications are \emph{degenerate} with our ignorance of the nuclear EoS. A dark core renders a star more compact, lowering both the radius and the tidal deformability at fixed mass, which mimics a softer nuclear EoS; an extended dark halo, by contrast, raises the gravitational mass and the tidal deformability and, hence, mimics a stiffer one~\cite{Giangrandi2023,Karkevandi2022}. Present mass--radius data are not sharp enough to separate the two effects from EoS uncertainty alone~\cite{Rutherford2023} yet. Breaking this degeneracy is among the central problems of the field, and it is precisely where self-consistent two-fluid modelling, as opposed to absorbing the dark matter into an effective single-fluid EoS, proves its worth~\cite{CiarcellutiSandin2011,EllisEtAl2018,IvanytskyiSagunLopes2020,Kain2021,DasMalikNayak2022,Giangrandi2023}.

Our point of entry is gravitational lensing, computed with the \GBT. The relevant technique is the optical-geometry method of Gibbons and Werner~\cite{GibbonsWerner2008}, in which the spatial paths of light on the equatorial plane are the geodesics of a two-dimensional Riemannian optical metric, and the deflection angle is evaluated by integrating the Gaussian curvature of that metric over the region the ray bounds. The construction has two features that make it well suited to the present problem. It is coordinate invariant, and it is evaluated over a domain lying \emph{outside} the light ray, which makes the result insensitive to the intricate boundary conditions at the surface of the central object and is sensitive instead to the geometry of the region the ray actually samples. It is this second feature that allows the surrounding dark-matter distribution to speak distinctly. The method has been applied widely to black holes, wormholes and horizonless compact objects, and has been extended to dispersive media, to finite source and observer distances, and to objects embedded in dark-matter halos~\cite{CrisnejoGallo2018,Ishihara2016,OnoAsada2019,Ovgun2019,LiuQiaoTao2024}.

It is worth being explicit about what is, and what is not, new here. The deflection caused by a compact object sitting in an \emph{imposed} dark-matter medium, whether a Hernquist, Dehnen or Burkert halo grafted onto a black-hole metric or a constant-density shell, is by now well-travelled ground~\cite{Ovgun2019,LiuQiaoTao2024}. Such calculations answer a different question: given a halo, how does light bend? They do not ask whether that halo is a configuration the matter would in fact adopt. The less-trodden and, we believe, more defensible step is to let the halo emerge as the solution of the coupled stellar-structure equations, so that the dark-matter fraction, the halo radius, and the external density profile are 
outputs of a single physical model, and only then to compute the deflection. The lensing observable is thereby tied to the same microphysics, namely the two equations of state and the astroparticle properties of the dark matter, that governs the mass, the radius, and the tidal deformability. It is this tethering that makes the observable useful as a degeneracy-breaking probe rather than a mere curiosity.

A second remark concerns the role of cooling that gave the original investigation its basis. The thermal state of the star enters here only insofar as it selects the configuration we model: an old, cold neutron star whose external baryonic pressure has long since fallen to negligible values, so that the external structure is set by the gravitating mass and by whatever dark-matter atmosphere extends beyond the baryonic surface. We make that assumption precise in \S~\ref{sec:twofluid}, rather than appealing to a higher-dimensional construction, and we set aside the electroweak-neutrino and ad-hoc plasma media of the preliminary study, whose gravitational lensing effect is both small and, more tellingly, not self-consistent with the stellar model~\cite{IvanytskyiSagunLopes2020}.

The paper is organised as follows. \S~\ref{sec:twofluid} sets out the two-fluid Tolman--Oppenheimer--Volkoff (TOV) system, distinguishes the dark core from the extended halo by means of Birkhoff's theorem, and specifies the external profile that feeds the lensing calculation. \S~\ref{sec:gbt} reviews the Gibbons--Werner method and fixes our conventions on the Schwarzschild test case. \S~\ref{sec:deflection} derives the weak deflection angle for the halo configuration, identifies the regime in which the halo leaves a signature, and specifies the numerical experiments required to quantify it, along with a raytracing experiment. \S~\ref{sec:discussion} discusses the result in the context of the EoS--dark-matter degeneracy and the prospects for observation, and \S~\ref{sec:conclusion} concludes. We work on the equatorial plane throughout, adopt a mostly-plus signature, and retain the factors of $G$ and $c$ explicitly in all principal results.

\section{A self-consistent two-fluid neutron star}
\label{sec:twofluid}

\subsection{The coupled TOV system}
\label{subsec:tov}

We model the star as two perfect fluids, a baryonic component (subscript $\mathrm{B}$) and a dark-matter component (subscript $\chi$), sharing a single static and spherically symmetric spacetime,
\begin{equation}
\dd s^{2} = -e^{2\Phi(r)}c^{2}\dd t^{2} + e^{2\lambda(r)}\dd r^{2}
           + r^{2}\qty(\dd\theta^{2} + \sin^{2}\theta\,\dd\varphi^{2}),
\label{eq:metric}
\end{equation}
in which the metric potentials $\Phi(r)$ and $\lambda(r)$ are dimensionless and
\begin{equation}
e^{-2\lambda(r)} = 1 - \frac{2G\,m(r)}{c^{2}r},
\qquad
m(r) = \mB(r) + \md(r)
\label{eq:massfunction}
\end{equation}
defines the (total) enclosed gravitational mass $m(r)$ as the sum of both the baryonic and dark-matter contributions. The two components interact only through gravity --- their stress-energy tensors are separately conserved,
$\nabla_{\mu}T^{\mu\nu}_{(\mathrm{B})}=0$ and
$\nabla_{\mu}T^{\mu\nu}_{(\chi)}=0$, 
with no direct exchange of energy or momentum. This is the standard multi-fluid generalization of the TOV equations~\cite{SandinCiarcelluti2009,LeungChuLin2011,TolosSchaffner2015}. Each fluid feeds the enclosed mass through its own energy density $\varepsilon_{i}$,
\begin{equation}
\dv{\mB}{r} = \frac{4\pi r^{2}}{c^{2}}\,\varepsilon_{\mathrm{B}},
\qquad
\dv{\md}{r} = \frac{4\pi r^{2}}{c^{2}}\,\varepsilon_{\chi},
\label{eq:dmdr}
\end{equation}
and each obeys hydrostatic balance against the \emph{common} metric potential,
\begin{equation}
\dv{p_{i}}{r} = -\bigl(\varepsilon_{i} + p_{i}\bigr)\dv{\Phi}{r},
\qquad i = \mathrm{B},\,\chi,
\label{eq:hydro}
\end{equation}
where the potential gradient is sourced by the \emph{total} enclosed mass and the \emph{total} pressure,
\begin{equation}
\dv{\Phi}{r} =
\frac{G\bigl[\,m(r) + 4\pi r^{3}(\pB + \pd)/c^{2}\,\bigr]}
     {c^{2} r\bigl[\,r - 2G\,m(r)/c^{2}\,\bigr]}.
\label{eq:dphidr}
\end{equation}
Eqs.~\eqref{eq:dmdr}--\eqref{eq:dphidr} are determined by two equations of state, $\pB(\varepsilon_{\mathrm{B}})$ and $\pd(\varepsilon_{\chi})$, and integrated outward from a pair of central densities $(\varepsilon_{\mathrm{B},c},\varepsilon_{\chi,c})$, or equivalently from a central baryonic density and a dark-matter mass fraction,
\begin{equation}
f_{\chi} \equiv \frac{\md(\Rd)}{\mB(\Rb)+\md(\Rd)} .
\label{eq:fchi}
\end{equation}
The two fluids in general reach vanishing pressure at different radii,
\begin{equation}
\pB(\Rb) = 0, \qquad \pd(\Rd) = 0,
\label{eq:radii}
\end{equation}
which define the baryonic radius $\Rb$ and the dark-matter radius $\Rd$.

For the baryonic sector we adopt a modern relativistic-mean-field EoS drawn from recent realistic dense-matter modelling~\cite{PanotopoulosLopes2017,Sagun2020}; for the dark sector we take a self-interacting (bosonic or fermionic) EoS of the kind used in recent two-fluid studies~\cite{Karkevandi2022,Giangrandi2023,IvanytskyiSagunLopes2020}.

 {Our numerical model consists of a baryonic core of mass $\MB=1.4\,\Msun$ and a diffuse dark-matter halo of outer radius $\Rd$ and mass fraction $f_{\chi}$, following the asymmetric-fermionic-dark-matter framework of Ivanytskyi, Sagun and Lopes~\cite{IvanytskyiSagunLopes2020}. The baryonic core is governed by a realistic, tabulated EoS, while the spatial extent of the dark sector is controlled by the dark-matter particle mass $m_{\mathrm{DM}}$: heavy particles form a compact centre that reduces the star's visible mass and radius, whereas light particles ($m_{\mathrm{DM}}\lesssim0.2$\,GeV) are pressure-supported to extend beyond the baryonic surface $\Rb$, creating an outer halo ($\Rd>\Rb$) that raises the star's visible gravitational mass.}
 
To ensure complete numerical reproducibility across all macroscopic structural calculations along with EoS profiles and lensing observables, the specific microphysical, and numerical parameters adopted in our calculations are explicitly summarized in Table~\ref{tab:parameters}. For the baryonic sector, we utilize canonical tabulated nuclear EoS models, specifically SLy4 and APR4, setting a baseline baryonic core mass of $M_{\rm B} = 1.4\,M_\odot$. For the dark matter halo component, we evaluate dark mass fractions across $f_\chi \in [0.05, 0.40]$, dark-matter particle masses $m_{\mathrm{DM}} \in [0.10, 0.20]\text{ GeV}$, and halo boundary radii $\Rd \in [15.0, 50.0]\text{ km}$, with the dark self-repulsion set by the interaction scale $y_v/m_v \in [10, 30]\text{ GeV}^{-1}$. The coupled two-fluid TOV equations and optical trajectory integrals are integrated numerically using a fine spatial grid step of $\Delta r = 1.0\text{ m}$.

\begin{table}[h!]
\centering
\caption{Microphysical and macroscopic parameters used in the numerical simulations.}
\label{tab:parameters}
\begin{tabular}{ll}
\hline\hline
\textbf{Parameter} & \textbf{Adopted values/model} \\
\hline
Baryonic EoS & SLy4, APR4 (tabulated)\\
Baryonic core mass, $M_{\rm B}$ & $1.4\,M_\odot$ \\
Standard core radius, $R_{\rm B}$ & $10.0\text{ km}$ (SLy4) \\
Standard core radius, $R_{\rm B}$ & $13.0\text{ km}$ (stiffened) \\
Dark matter mass fraction, $f_\chi$ & $0.05 - 0.40$,  step $0.012$\\
Halo boundary radius, $R_{\chi}$ & $15.0 - 50.0\text{ km}$ \\
Dark matter particle mass, $m_{\mathrm{DM}}$ & $0.10 - 0.20\text{ GeV}$ \\
Dark matter interaction, $\frac{y_v}{m_v}$ & $10 - 30\text{ GeV}^{-1}$ (scale) \\
Integration spatial step, $\Delta r$ & $1.0\text{ m}$ \\
\hline\hline
\end{tabular}
\end{table}

\subsection{Dark core versus extended halo}
\label{subsec:corehalo}

The relation between $\Rd$ and $\Rb$ sorts the solutions into two physically distinct classes, and the distinction proves decisive for lensing.

\paragraph*{Confined dark core ($\Rd<\Rb$).}
All of the dark matter lies within the baryonic surface. For $r>\Rb$ the
spacetime is vacuum, and Birkhoff's theorem fixes the exterior to be exactly Schwarzschild with the total gravitational mass $\Mtot=\mB(\Rb)+\md(\Rb)$. A light ray passing outside the star therefore feels the dark matter \emph{only} through that total mass and the compactness it implies; the admixture produces no lensing feature distinguishable from that of a more massive, or more compact, ordinary star. The core is, in this precise sense, invisible to the geometric probe, however much it may matter for the internal structure.

\paragraph*{Extended dark halo ($\Rd>\Rb$).}
The dark matter reaches beyond the baryonic surface, so that the shell $\Rb<r<\Rd$ is filled with a dilute dark-matter atmosphere and only the region $r>\Rd$ is true vacuum. By Birkhoff's theorem the metric is Schwarzschild with the full mass $\Mtot=\mB(\Rb)+\md(\Rd)$ \emph{only} for $r>\Rd$; within the halo the enclosed mass $m(r)$ is still growing, $g_{rr}$ has not yet reached its Schwarzschild form, and $\Phi(r)$ is not the vacuum logarithm. A ray with impact parameter $b<\Rd$ samples this non-Schwarzschild atmosphere, and its deflection records the halo profile rather than merely the total mass. This is the configuration we model, and the regime in which the Gibbons--Werner construction isolates the surrounding density.

Whether a given star develops a halo or a core is set by the dark-matter
microphysics. The robust trend across the two-fluid literature is that light particles (sub-GeV) and strong self-repulsion favour extended, dilute haloes, whereas heavier and weakly interacting particles favour compact cores~\cite{LeungChuLin2011,NelsonReddyZhou2019,Karkevandi2022,Giangrandi2023,Rutherford2023}; a larger dark-matter fraction $f_{\chi}$ amplifies whichever regime applies. We therefore restrict attention to the halo-forming corner of parameter space, and treat $(f_{\chi},\Rd)$ as the structural quantities the lensing observable will constrain.

\subsection{The external profile that feeds the lensing}
\label{subsec:profile}

For the deflection calculation we require the metric functions in the region the ray traverses, $r>\Rb$. There the baryonic contribution is frozen, $\mB(r)=\mB(\Rb)\equiv\MB$, while dark-matter mass continues to accumulate,
\begin{equation}
\begin{gathered}
m(r) = \MB + \md(r),\\[3pt]
\md(r) = 4\pi\!\int_{\Rb}^{r} \rhod(r')\, r'^{2}\,\dd r' ,
\end{gathered}
\label{eq:halomass}
\end{equation}
for $\Rb \le r \le \Rd$, and saturating at $\md(\Rd)$ for $r\ge\Rd$. Here, $\rhod(r)=\varepsilon_{\chi}(r)/c^{2}$ is the dark-matter \emph{mass} density, which falls monotonically and smoothly from $\rhod(\Rb)$ to zero at $\Rd$. There is no closed-form expression for $\rhod(r)$; it is the numerical output of Eqs.~\eqref{eq:dmdr}--\eqref{eq:dphidr}. For the weak-field estimates below, it is convenient to characterise the halo by its mass fraction $f_{\chi}$, its radius $\Rd$, and the normalised enclosed-mass function $\mu(r)\equiv m(r)/\Mtot$, which rises monotonically from $\mu(\Rb)=\MB/\Mtot=1-f_{\chi}$ at the baryonic surface to $\mu(\Rd)=1$ at the edge of the halo.

\section{Optical geometry and the Gauss--Bonnet deflection angle}
\label{sec:gbt}

\subsection{The Gibbons--Werner construction}
\label{subsec:gw}

Light follows null geodesics, $\dd s^{2}=0$. On the equatorial plane $\theta=\pi/2$, the metric~\eqref{eq:metric} furnishes a Fermat (optical) metric for the spatial path,
\begin{equation}
c^{2}\dd t^{2}
= \frac{e^{2\lambda(r)}}{e^{2\Phi(r)}}\,\dd r^{2}
+ \frac{r^{2}}{e^{2\Phi(r)}}\,\dd\varphi^{2}
\equiv g^{\mathrm{opt}}_{ij}\,\dd x^{i}\dd x^{j},
\label{eq:optical}
\end{equation}
a two-dimensional Riemannian metric whose geodesics are the spatial light rays. The Gibbons--Werner method applies the \GBT\ to a domain $\mathcal{D}$ of this optical surface bounded by the ray and by a circular arc at large radius~\cite{GibbonsWerner2008}. Writing the Gauss--Bonnet identity as
\begin{equation}
\iint_{\mathcal{D}} \mathcal{K}\,\dd S + \oint_{\partial\mathcal{D}} \kappa_{g}\,\dd\ell + \sum_{j}\theta_{j} = 2\pi\,\chi_{\mathrm{E}}(\mathcal{D}),
\label{eq:gbt}
\end{equation}
where $\chi_{\mathrm{E}}(\mathcal{D})$ is the Euler characteristic (equal to unity for our disk-topology domain), the vanishing of the geo-desic curvature $\kappa_{g}$ along the ray together with the asymptotic flatness of the optical metric collapses Eq.~\eqref{eq:gbt} into the deflection angle 
\begin{equation}
\hat\alpha = -\iint_{\mathcal{D}} \mathcal{K}\,\dd S ,
\label{eq:alpha_master}
\end{equation}
with $\mathcal{K}$ the Gaussian curvature of the optical metric~\eqref{eq:optical} and $\dd S = \sqrt{g^{\mathrm{opt}}}\,\dd r\,\dd\varphi$ its area element. The deflection is thus, a curvature integral over the region the ray bounds, a partly topological statement, and the source of the method's robustness to the internal boundary conditions of the lens.

Two qualifications from the recent literature bear on our use of Eq.~\eqref{eq:alpha_master}. First, strict asymptotic flatness holds only for $r>\Rd$, where the metric is vacuum Schwarzschild; because the halo has finite extent, the global spacetime \emph{is} asymptotically flat and Eq.~\eqref{eq:alpha_master} applies, with the integrand $\mathcal{K}$ taken from the two-fluid optical metric inside $\Rd$ and from Schwarzschild outside. Second, should one wish to place source and observer at finite distance, as is appropriate for a Galactic neutron star, the finite-distance generalisation of the \GBT\ due to Ishihara, Suzuki, Ono and Asada should be employed~\cite{Ishihara2016,OnoAsada2019}; it reduces to Eq.~\eqref{eq:alpha_master} as the inverse source and observer distances tend to zero. We quote the asymptotic result as the principal one, and indicate the finite-distance correction where it matters.

\subsection{Conventions: the Schwarzschild test case}
\label{subsec:schwarzschild}

For a vacuum Schwarzschild lens, $e^{2\Phi}=e^{-2\lambda}=f(r)=1-2GM/(c^{2}r)$, the optical metric~\eqref{eq:optical} reduces to
\begin{equation} 
c^{2}\dd t^{2} = \frac{\dd r^{2}}{f(r)^{2}} + \frac{r^{2}}{f(r)}\,\dd\varphi^{2},
\label{eq:schw_opt}
\end{equation}
whose Gaussian curvature is
\begin{equation}
\mathcal{K} = -\frac{2GM}{c^{2}r^{3}}\qty(1 - \frac{3GM}{2c^{2}r})
+ \mathcal{O}\!\qty(\frac{G^{3}M^{3}}{c^{6}r^{6}}),
\label{eq:schw_K}
\end{equation}
negative near the lens. Inserting Eq.~\eqref{eq:schw_K} into the master formula~\eqref{eq:alpha_master}, with the leading-order straight-line boundary $r=b/\sin\varphi$, recovers the Einstein result; the systematic weak-field expansion in the impact parameter $b$ reads~\cite{KeetonPetters2005}
\begin{equation}
\begin{aligned}
\hat\alpha_{\mathrm{S}}
={}& \frac{4GM}{c^{2}b}
+ \frac{15\pi}{4}\qty(\frac{GM}{c^{2}b})^{2}\\
&+ \frac{128}{3}\qty(\frac{GM}{c^{2}b})^{3}
+ \mathcal{O}\!\qty(\frac{GM}{c^{2}b})^{4}.
\end{aligned}
\label{eq:schw_series}
\end{equation}
We have verified the curvature~\eqref{eq:schw_K} and the first three coefficients of Eq.~\eqref{eq:schw_series} by symbolic computation; they are the standard Schwarzschild impact-parameter coefficients $\{4,\,15\pi/4,\,128/3\}$.  {The expansion is written in the gauge-independent impact parameter $b$; had it instead been organised in the coordinate-dependent closest-approach radius $r_{0}$, the higher-order coefficients would differ. We therefore use $b$ consistently here and in the lensing analysis of \S~\ref{sec:deflection}.}

The leading term $4GM/(c^{2}b)$ is the only one that survives for a ray confined to the vacuum exterior, $b>\Rd$, with $M=\Mtot$; the halo signature we seek is a departure from this behaviour for $b<\Rd$.

\section{Weak deflection by the extended dark-matter halo}
\label{sec:deflection}

\subsection{The two regimes}
\label{subsec:tworegimes}

\begin{figure}[htbp]
    \centering
    \includegraphics[width=0.48\textwidth]{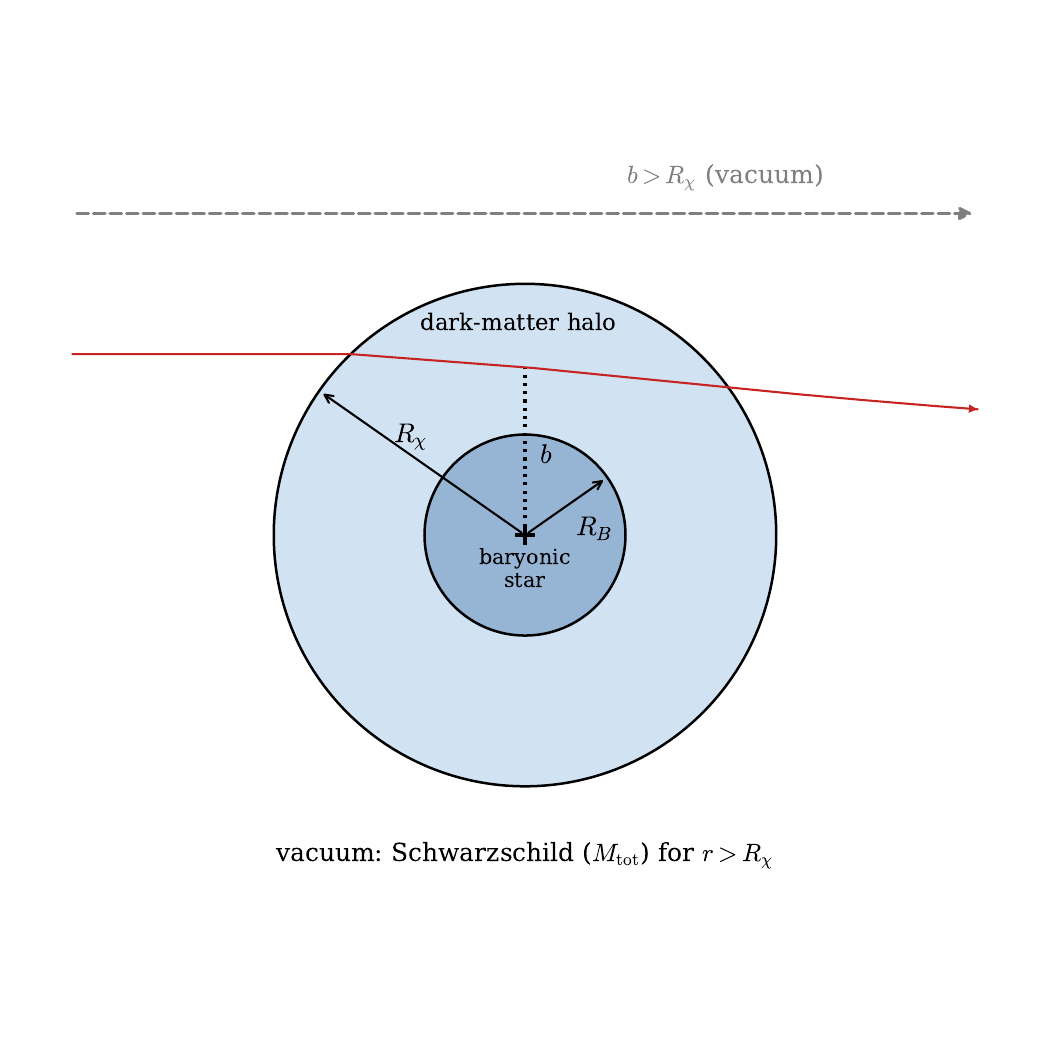}
    \caption{Geometry of the two configurations and the lensing set-up. The baryonic star ($r<\Rb$) is embedded in a dark-matter halo reaching to $\Rd$, beyond which the spacetime is vacuum and, by Birkhoff's theorem, Schwarzschild with the total mass $\Mtot$. A ray with impact parameter $b<\Rd$ (red) threads the halo, so its deflection records the enclosed profile; a ray with $b>\Rd$ remains in vacuum and feels only $\Mtot$. 
    }
    \label{fig:schematics}
\end{figure}

The deflection of a ray of impact parameter $b$ separates cleanly according to whether the ray enters the halo. Fig.~\ref{fig:schematics} depicts the schematic of our setup.

\paragraph*{Grazing the vacuum exterior ($b>\Rd$).}
The ray never leaves the Schwarzschild region. By Birkhoff's theorem it is deflected exactly as by a point mass $\Mtot$,
\begin{equation}
\hat\alpha(b) = \frac{4G\Mtot}{c^{2}b}
+ \frac{15\pi}{4}\qty(\frac{G\Mtot}{c^{2}b})^{2} + \cdots,
\qquad b>\Rd .
\label{eq:alpha_outside}
\end{equation}
Here, the dark matter is detectable only as additional mass, degenerate with a heavier baryonic star; no geometric signature survives.

\paragraph*{Traversing the halo ($b<\Rd$).}
The ray now samples the non-Schwarzschild shell, and the curvature integral \eqref{eq:alpha_master} runs over a domain in which $\mathcal{K}$ is built from the two-fluid optical metric~\eqref{eq:optical} with the growing mass function
$m(r)$ of Eq.~\eqref{eq:halomass}. To leading (weak-field) order, the Gaussian curvature of a static, spherically symmetric optical metric is
\begin{equation}
\mathcal{K}(r) \simeq \Phi''(r) + \frac{\Phi'(r)}{r} + \frac{G}{c^{2}}\!\left[\frac{1}{r^{2}}\,m'(r) - \frac{m(r)}{r^{3}}\right] + \cdots ,
\label{eq:K_general}
\end{equation}
in which a prime denotes $\dd/\dd r$ and both the temporal potential $\Phi$ and the spatial mass function $m(r)$ contribute. Eq.~\eqref{eq:K_general} reduces correctly to the Schwarzschild curvature~\eqref{eq:schw_K} in vacuum, where $m(r)=M$ is constant and $\Phi(r)=-GM/(c^{2}r)$ at leading order, the two pieces then combining to give $-2GM/(c^{2}r^{3})$. Carrying out the integral in Eq.~\eqref{eq:alpha_master} along the straight-line ray $r=b/\sin\varphi$, and using that in the weak field the temporal and spatial metric terms contribute equally (the origin of the familiar factor of two over the Newtonian value), yields the central analytic result of this paper: a deflection governed by \emph{the mass which the ray actually encircles} rather than by the total mass,
\begin{equation}
\hat\alpha(b) \simeq \frac{4G}{c^{2}b}\,\Mcyl(b) + \hat\alpha_{p}(b) + \cdots , \qquad b<\Rd ,
\label{eq:alpha_halo}
\end{equation}
where $\Mcyl(b)$ is the mass contained in the cylinder of radius $b$ threaded by the ray, the standard projected-mass form of the weak-lensing deflection~\cite{SchneiderEhlersFalco1992,BartelmannSchneider2001},
\begin{equation}
\Mcyl(b) = \MB + 4\pi\!\int_{\Rb}^{\Rd}\rhod(r)\,\mathcal{W}(r,b)\, r^{2}\,\dd r ,
\label{eq:Mcyl}
\end{equation}
and the projection weight, which maps a spherical shell of radius $r$ onto the cylinder of radius $b$, is
\begin{equation}
\mathcal{W}(r,b) = 
\begin{cases}
1, & r \le b,\\[4pt]
\displaystyle 1 - \sqrt{1 - b^{2}/r^{2}}, & r > b .
\end{cases}
\label{eq:weight}
\end{equation}
The residual term $\hat\alpha_{p}(b)$ collects the subdominant contribution of the dark-matter pressure through $\Phi'(r)$ in Eq.~\eqref{eq:dphidr}, negligible for a dilute, non-relativistic halo. We have confirmed Eqs.~\eqref{eq:alpha_halo}--\eqref{eq:weight} numerically: for a representative halo profile, the cylinder mass~\eqref{eq:Mcyl} agrees with an independent surface-density projection to one part in $10^{7}$, and reproduces the point-mass
limit $\hat\alpha\to 4G\Mtot/(c^{2}b)$ as $b$ grows beyond $\Rd$.

Eq.~\eqref{eq:alpha_halo} renders quantitative the qualitative claim of the introduction. For $b<\Rd$ the ray encircles only the mass interior to its own cylinder, which is strictly less than $\Mtot$ because part of the halo lies beyond $b$. The deflection is correspondingly \emph{suppressed} relative to the point-mass value $4G\Mtot/(c^{2}b)$, the suppression deepening as $b$ decreases into the halo and as the dark-matter fraction $f_{\chi}$ and the halo radius $\Rd$ grow. A diffuse halo redistributes deflecting mass to large radii, flattening the $\hat\alpha(b)$ curve at small $b$ in a manner no point mass can imitate. This shape, and not the overall normalisation, which remains degenerate with mass, is the observable signature.

\subsection{Numerical experiments}
\label{subsec:numerics}

\begin{figure}[t]
\centering
\includegraphics[width=0.48\textwidth]{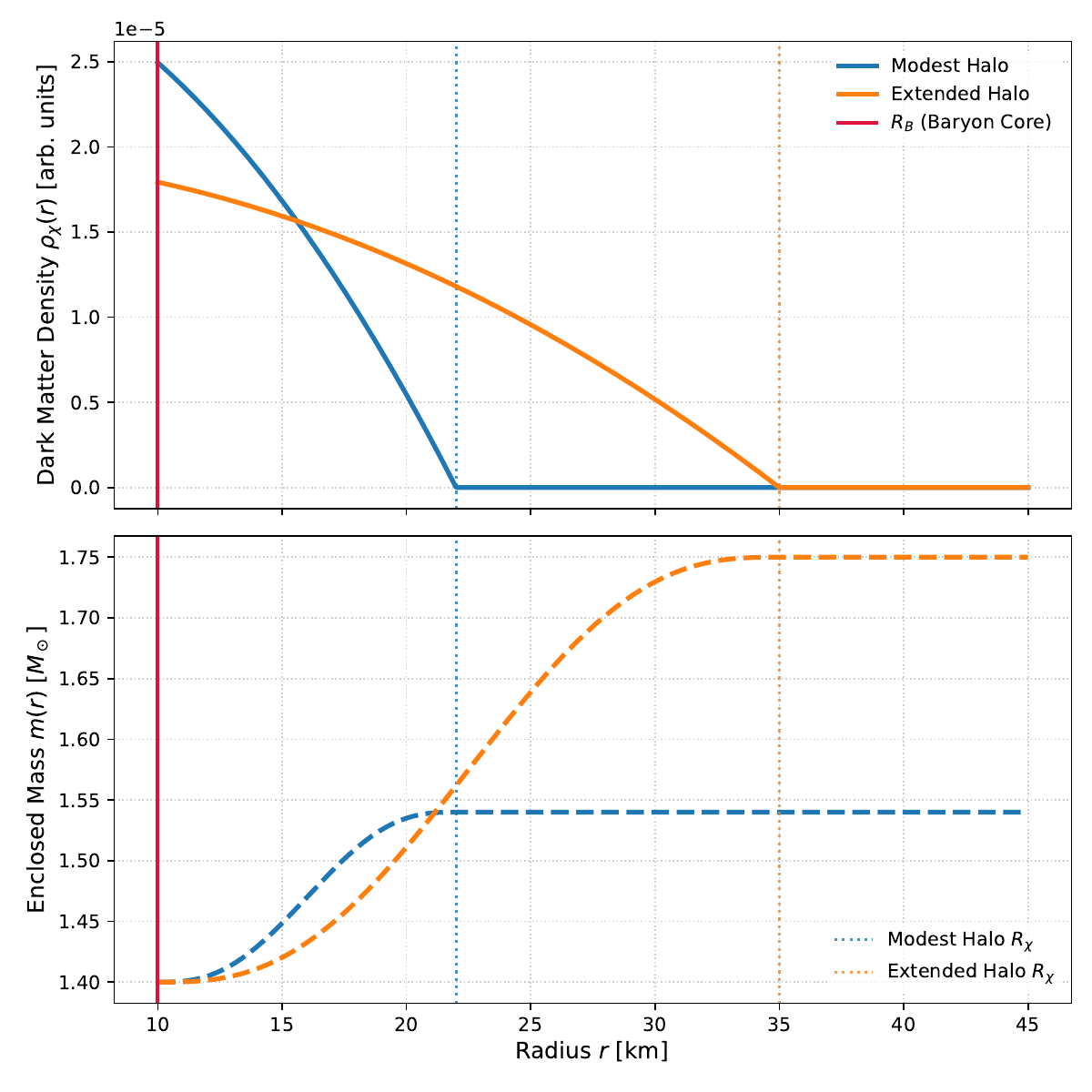}
\caption{ {Self-consistent external profiles feeding the lensing calculation. Dark-matter density $\rhod(r)$ (upper panel) and enclosed mass $m(r)$ (lower panel) for a modest halo ($f_{\chi}=0.10$, $\Rd=22$\,km) and an extended halo ($f_{\chi}=0.25$, $\Rd=35$\,km), with $\Rb=10$\,km. The density falls smoothly to zero at $\Rd$, while $m(r)$ rises from $\MB$ at $\Rb$ and saturates at $\Mtot$ for $r\ge\Rd$.}}
\label{fig:profile}
\end{figure}

\begin{figure}[t]
\centering
\includegraphics[width=0.48\textwidth]{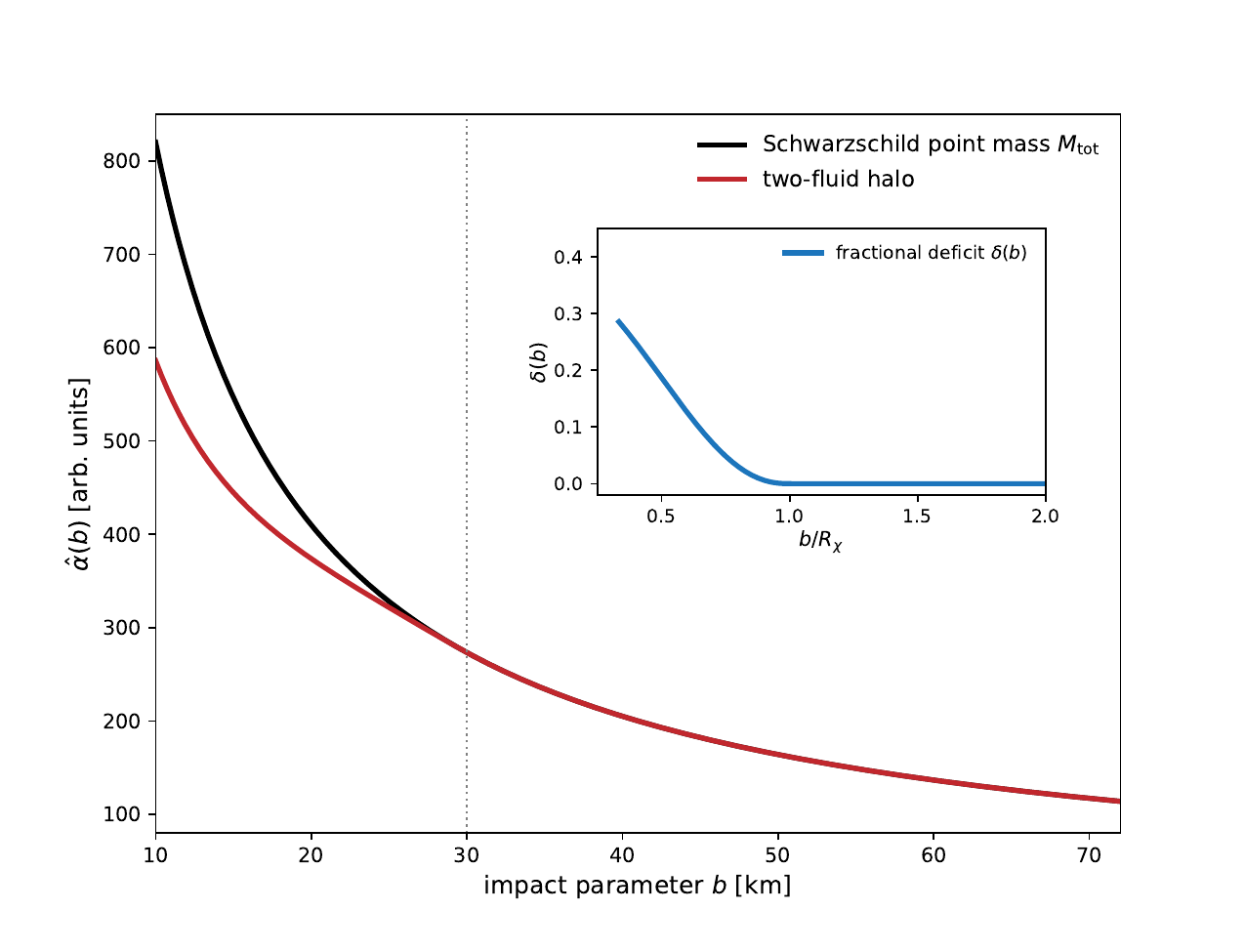}
\caption{ {The halo signature, for a representative star ($f_{\chi}=0.35$, $\Rd=30$\,km, $\Rb=10$\,km). Weak deflection angle $\hat\alpha(b)$ (arbitrary units): the Schwarzschild point mass with $M=\Mtot$ (black) and the two-fluid halo (red), which falls below the point mass for $b<\Rd$ and rejoins it at $b=\Rd$. Inset: the fractional deficit $\delta(b)=1-\Mcyl(b)/\Mtot$ versus $b/\Rd$, largest at the baryonic surface and decreasing monotonically to zero at $b=\Rd$.}}
\label{fig:deflection}
\end{figure}

\begin{figure}[t]
\centering
\includegraphics[width=0.48\textwidth]{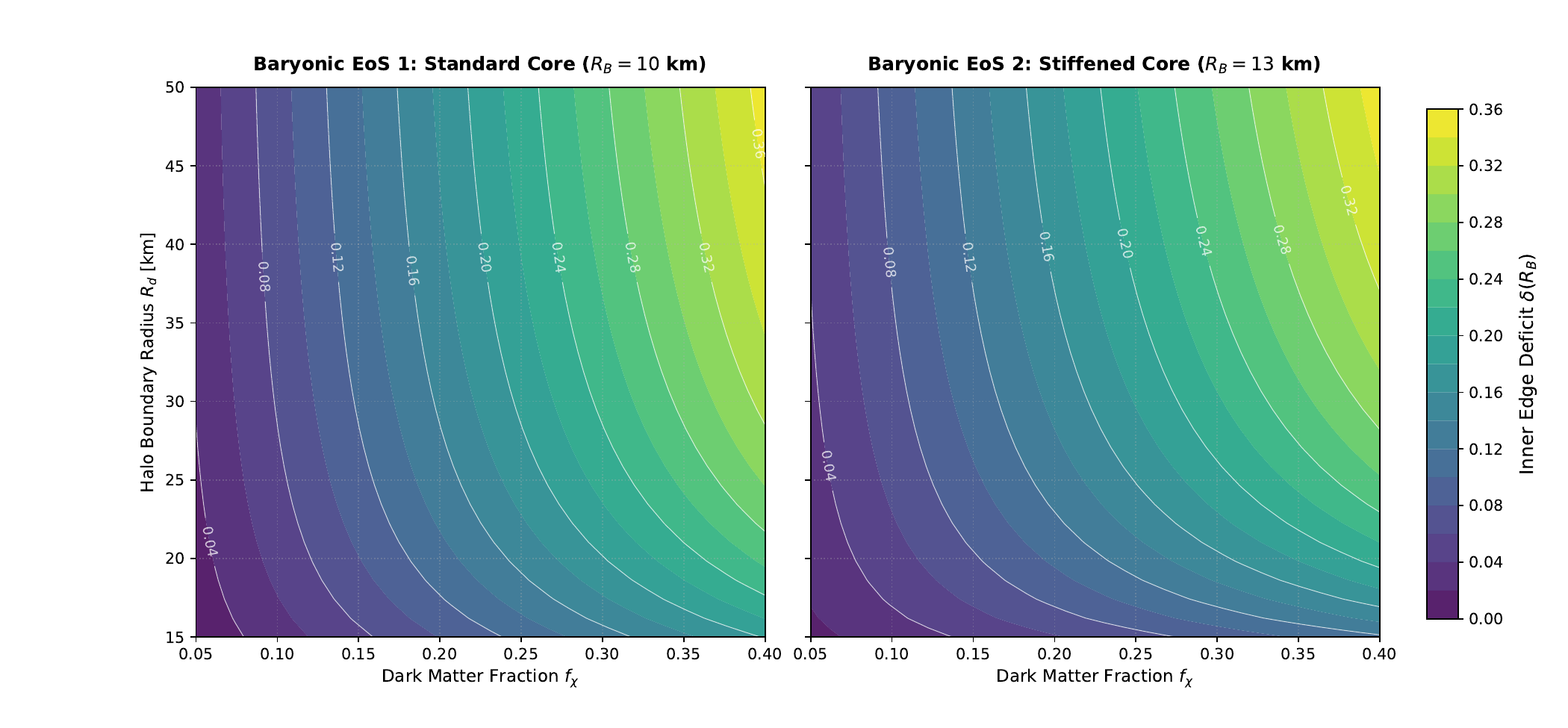}
\caption{ {Lensing as a handle on the EoS--dark-matter degeneracy. Inner-edge deficit $\delta(\Rb)$ across the $(f_{\chi},\Rd)$ plane, for a standard baryonic core ($\Rb=10$\,km, left) and a stiffer one ($\Rb=13$\,km, right). The contours tilt with $\Rd$ and separate the two equations of state, so that the deficit responds to both the amount and the spatial extent of the dark matter rather than to the total mass alone.}}
\label{fig:degeneracy}
\end{figure}

The analytic result~\eqref{eq:alpha_halo} fixes the structure of the signal; its magnitude comes from the self-consistent profiles.  {We carry out four runs; their principal outputs are Figs.~\ref{fig:profile}--\ref{fig:degeneracy}.}
\begin{enumerate}
\item \emph{Stellar sequences.}  {We integrate} Eqs.~\eqref{eq:dmdr}--\eqref{eq:dphidr} for the chosen EoS pair over a grid in central baryonic density and dark-matter fraction $f_{\chi}$, identify the halo-forming region ($\Rd>\Rb$) and tabulate $(\MB,\md,\Rb,\Rd,\Mtot)$ along representative sequences.

\item \emph{External profiles.}  {For a modest and an extended halo we output the density $\rhod(r)$ and enclosed mass $m(r)$} on $\Rb<r<\Rd$ (Fig.~\ref{fig:profile}).

\item \emph{Deflection curves.}  {We evaluate} $\hat\alpha(b)$ from Eqs.~\eqref{eq:alpha_master} and~\eqref{eq:alpha_halo} across $b$, spanning the halo ($b<\Rd$) and the vacuum exterior ($b>\Rd$), and overlay the Schwarzschild point-mass curve with $M=\Mtot$ (Fig.~\ref{fig:deflection}). The diagnostic is the fractional deficit
$\delta(b)\equiv 1-\hat\alpha(b)\,c^{2}b/(4G\Mtot)$ as a function of $b/\Rd$.

\item \emph{Parameter dependence.}  {We map the inner-edge def-icit $\delta(\Rb)$, which is where $\delta(b)$ is largest,} over the $(f_{\chi},\Rd)$ plane and against the baryonic EoS, exhibiting the degeneracy-breaking power discussed in \S~\ref{sec:discussion} (Fig.~\ref{fig:degeneracy}).
\end{enumerate}

 {Figure~\ref{fig:profile} contrasts a \emph{modest} halo ($f_{\chi}=0.10$, $\Rd=22.0$\,km) with an \emph{extended} halo ($f_{\chi}=0.25$, $\Rd=35.0$\,km), showing how the dark-matter density $\rhod(r)$ and the enclosed mass $m(r)$ evolve across the halo boundary in the light-particle regime; the enclosed mass rises from $\MB$ at $\Rb$ and saturates at $\Mtot$ at $\Rd$. For a representative star ($f_{\chi}=0.35$, $\Rd=30.0$\,km, $\Rb=10.0$\,km), Fig.~\ref{fig:deflection} shows the weak-field deflection $\hat\alpha(b)$ falling increasingly below the Schwarzschild point mass of equal total mass as the ray penetrates the halo ($b<\Rd$), with a fractional deficit $\delta(b)$ that is largest at the baryonic surface and vanishes at $b=\Rd$. Mapping this inner-edge deficit $\delta(\Rb)$ over $f_{\chi}\in[0.05,0.40]$ and $\Rd\in[15.0,50.0]$\,km (Fig.~\ref{fig:degeneracy}) yields contours that tilt with $\Rd$ and separate the two baryonic equations of state ($\Rb=10.0$\,km against a stiffer $\Rb=13.0$\,km), so that the deficit encodes both the amount and the spatial extent of the dark matter, not the total mass alone.}

\subsubsection*{Ray-tracing experiment}

 {To visualise the analytic result Eq.~\eqref{eq:alpha_halo} in the image plane, we carry out a two-dimensional ray-tracing experiment, tracing rays backward from the observer plane through the lens.}

\begin{figure*}[t!]
\centering
\includegraphics[width=0.98\textwidth]{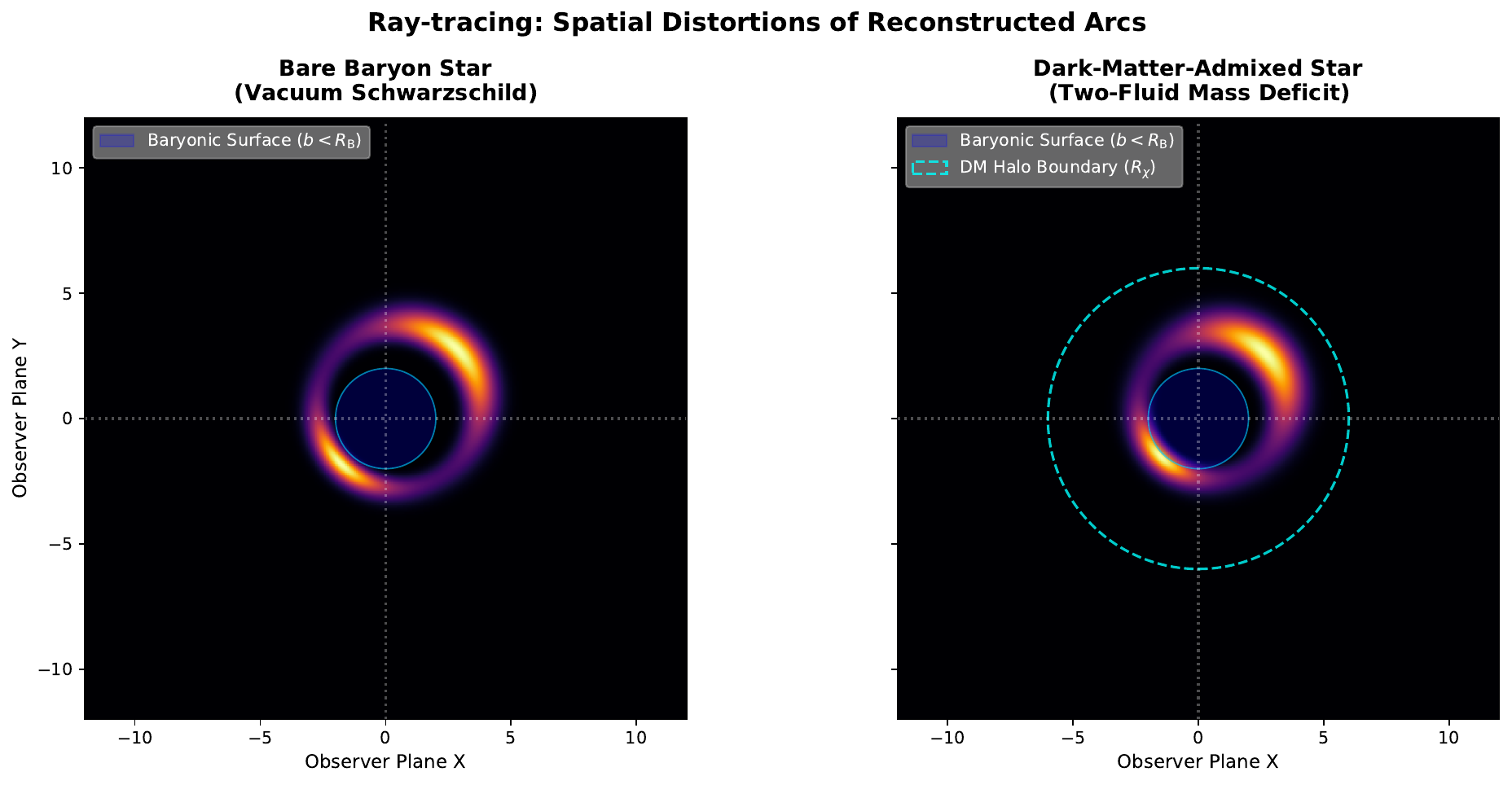}
\caption{A two-dimensional lens-plane reconstruction of an off-centre background source, comparing a vacuum Schwarzschild baseline (left) to our self-consistent two-fluid dark-matter--admixed star (right). Within the baryonic surface ($b<\Rb$) passing rays are extinguished. In the admixed scenario, rays piercing the extended halo boundary ($b < \Rd$, dashed cyan circle) experience a suppressed deflection field due to the cylinder mass deficit $\Mcyl(b)$.  {Both panels share the same $\Mtot$ and background source, so the difference between them, chiefly a slight outward shift of the counter-image (lower left), is a profile effect and not a mass effect.} 
}
\label{fig:raytracing}
\end{figure*}

Figure~\ref{fig:raytracing} displays a reconstructed observer lens plane for an extended, unlensed circular background source passing slightly off-centre behind the central lens. We contrast a conventional pure baryon star against our two-fluid dark-matter-admixed star of identical total mass $\Mtot$. The offset to the unlensed background source breaks the azimuthal degeneracy; without it the image would be a symmetric Einstein ring, every point of which shares one impact parameter and so samples the deflection at a single radius. The offset instead forces the rays to span a continuous range of impact parameters, across the vacuum exterior ($b>\Rd$) and the non-Schwarzschild shell ($\Rb<b<\Rd$) alike, which separates the total-mass normalization $\Mtot$ from the underlying spatial mass gradient and isolates the non-point-mass signature of $\Mcyl(b)$ in the differential shear of the reconstructed arcs.

The difference between the two panels illustrates the degeneracy-breaking content of the \GBT\ deflection. Both configurations show a central silhouette for $b<\Rb$, where rays intersect the baryonic surface and are extinguished. For $\Rb<b<\Rd$, however, the trajectories sample the non-Schwarzschild shell: because the halo redistributes the deflecting mass to larger radii, the cylinder mass $\Mcyl(b)$ is strictly less than $\Mtot$ and the deflection is suppressed as in Eq.~\eqref{eq:alpha_halo}.  {In the admixed panel the counter-image is accordingly displaced outward and tangentially stretched relative to the vacuum case; since the two panels share the same total mass and source, this difference is due to the profile alone.}

To quantitatively evaluate the observational scale of these deflection departures, we consider an isolated dark-matter--admixed neutron star at a fiducial distance of $d = 100\text{ pc}$. For a $1.4\,M_\odot$ system ($r_g \equiv GM_{\rm tot}/c^2 \approx 2.067\text{ km}$):
 \begin{itemize}
 \item the characteristic gravitational mass angle is $\theta_g \equiv r_g / d \approx 0.138\text{ nas}$ which is $1.38 \times 10^{-4}\,\mu\text{as}$,
 \item the physical subtension of the halo boundary at $R_{\chi} = 30\text{ km}$ which corresponds to $\theta_{\chi} \equiv R_{\chi}/d \approx 2.01\text{ nas}$, and
 \item baryonic surface ($R_{\rm B} = 10\text{ km}$) on the sky is $\theta_{\rm B} \equiv R_{\rm B}/d \approx 0.668\text{ nas}$. 
 \end{itemize}

 {We stress that the reconstruction lives at the scale of the gravitational radius. For any realistic neutron-star lens the rays that sample the halo lie deep inside the Einstein radius, in the strongly demagnified central image, so the figure is an illustration of how the profile reshapes the deflection field rather than a directly resolvable observable. Its purpose is to make visible, in the image plane, the same statement carried by Fig.~\ref{fig:deflection}: at fixed total mass, redistributing the mass into a halo changes the images, so the signature is a property of the profile and not of the mass.}

\section{Discussion}
\label{sec:discussion}

The value of Eq.~\eqref{eq:alpha_halo} is that it ties a lensing observable to the very two-fluid model that fixes the star's mass, radius and tidal deformability. The degeneracy that frustrates the structural observables, with a dark core softening and a dark halo stiffening the \emph{apparent} nuclear EoS~\cite{Giangrandi2023,Karkevandi2022,Rutherford2023}, enters the lensing problem in a different guise. The mass-like part of the deflection, Eq.~\eqref{eq:alpha_outside}, is indeed degenerate: a halo star and a heavier baryonic star bend distant light identically. The \emph{shape} of $\hat\alpha(b)$ for $b<\Rd$, however, encoded in Eqs.~\eqref{eq:alpha_halo} and~\eqref{eq:Mcyl}, depends on \emph{where} the mass resides, and a diffuse dark halo places mass at radii no plausible baryonic configuration can reach. In principle, then, a measurement that resolves the deflection at impact parameters comparable to the halo radius carries information the global mass and radius do not, and may help to separate genuine dark-matter admixture from EoS uncertainty, which is the central aim of the self-consistent
programme~\cite{IvanytskyiSagunLopes2020,Giangrandi2023}.

We are candid about the observational reach. The halo radii produced by two-fluid models are at most a few tens of kilometres, and the corresponding angular scales for any realistic neutron-star distance lie far below current direct-imaging resolution. The result is therefore best read as a theoretical diagnostic: a coordinate-invariant statement of how, and by how much, a self-consistent halo modifies light bending, and a demonstration that the modification is a profile effect rather than a mass effect.  {Its concrete value is as an independent, geometric cross-check on the halo configurations inferred from tidal deformability in gravitational-wave events~\cite{Karkevandi2022,Giangrandi2023,DasMalikNayak2022}: two very different probes of one and the same self-consistent model.}

Additionally, in terms of observational implications and feasibility, since compact object ray-tracing spans large intrinsic deflection angles ($\hat{\alpha} \sim 0.28\text{ rad} \approx 16^\circ$ at $b = \Rd$) and the weak-field deflection scales linearly with enclosed mass i.e., $\hat{\alpha}(b) \propto M(b)/b$, a fractional mass deficit $\delta(b)$ induces a corresponding first-order shift in the effective impact parameter, $\Delta b \approx \delta(b) \cdot b$ --- a sky-plane displacement $\Delta\theta(b) \equiv \Delta b / d \approx \delta(b) \cdot \theta(b)$ when projected to the observer plane. For an impact parameter probing the inner halo edge $b \approx \Rb$, a peak fractional mass deficit of $\delta(\Rb) = 0.35$ produces an apparent sky-plane displacement of $\Delta\theta \approx 0.23\text{ nas}$ ($2.3 \times 10^{-4}\,\mu\text{as}$) for $d = 100\text{ pc}$, scaling to $\Delta\theta \approx 2.3\text{ nas}$ for a system nearby at $d = 10\text{ pc}$.

Comparing these physical scales to modern observational limits where sub-millimeter VLBI (Event Horizon Telescope \cite{EHT2019_paperI}) achieves resolutions of $\sim 20\,\mu\text{as}$ and optical interferometry (VLTI/GRAVITY \cite{GRAVITYcollab2017,GRAVITYcollab2018}) reaches centroiding limits of $\sim 10\,\mu\text{as}$, it can be shown that the angular radius of the halo itself ($\theta_{\chi} \approx 2.01\text{ nas}$) is roughly four orders of magnitude below current direct-imaging thresholds, while the differential image displacement ($\Delta\theta \approx 0.23\text{ nas}$) lies nearly five orders of magnitude below present resolution. Consequently, the diagnostic signature of $\hat{\alpha}(b)$ is best probed not by means of direct shadow resolution, but through high-precision astrometric microlensing events (where the lens Einstein radius $\theta_E = \sqrt{4 r_g / d} \approx 10.68\text{ mas}$ is well within reach of Gaia-era astrometry \cite{Paczynski1996,Gaia2016,Sahu2022,Lam2022}) or distinct pulse-arrival time-delay signatures ($\Delta t \sim \int \Phi\,\mathrm{d}t$) in lensed pulsar systems \cite{Shapiro1964,LorimerKramer2004}.

We close by noting what we have deliberately avoided. We have not inserted a phenomenological halo profile and tuned its parameters to maximize the signal;
the halo here is whatever the coupled equations produce. 
We have not invoked higher-dimensional or modified-gravity constructions to justify the external metric: the assumption that an old, cold star carries negligible external baryonic pressure suffices, and is stated as such in \S~\ref{sec:twofluid}. The result stands or falls on the self-consistency of the two-fluid model and the coordinate invariance of the \GBT.

\section{Conclusions}
\label{sec:conclusion}

We have set out, and reduced to a single computable expression, the weak gravitational deflection of light by a neutron star carrying a self-consistent dark-matter halo. The physical content is threefold. A confined dark core is invisible to lensing, by Birkhoff's theorem. An extended dark halo, through which the ray passes, suppresses the deflection at impact parameters below the halo radius relative to the point-mass prediction, by an amount fixed by the dark-matter fraction and the halo extent, Eqs.~\eqref{eq:alpha_halo} and~\eqref{eq:Mcyl}. And because the halo is the solution of the coupled two-fluid TOV system rather than a medium inserted by hand, the lensing observable is tethered to the same microphysics that governs the star's mass, radius and tidal deformability, which is what makes the \emph{shape} of the deflection curve a candidate for breaking the degeneracy between dark matter and the nuclear EoS.

 {The numerical experiments of \S~\ref{subsec:numerics} bear this out. The fractional deficit $\delta(b)$ is largest at the baryonic surface, decreases monotonically and vanishes at $b=\Rd$ (Fig.~\ref{fig:deflection}), while its inner-edge value $\delta(\Rb)$ traces contours that tilt with the halo radius and separate a stiffer from a softer baryonic equation of state (Fig.~\ref{fig:degeneracy}). The original contribution of this work is to tie a coordinate-invariant lensing diagnostic to a self-consistent two-fluid stellar model, so that the deflection reflects the same microphysics as the mass, the radius and the tidal deformability, and to identify the \emph{shape} of the deficit, rather than its normalisation, as the quantity that separates an extended dark-matter halo from a stiffening of the nuclear equation of state. Because it rests on the same model as the tidal deformability, this geometric diagnostic provides an independent cross-check on dark-matter admixture of the kind inferred from gravitational-wave observations.}

Finally, we address the scope of our spherically symmetric spacetime approximation. Real neutron stars rotate, breaking spherical symmetry and introducing spin angular momentum $J$ and a rotation-induced mass quadrupole moment $Q$. In this realistic setting, the vacuum exterior ($r > \Rd$) transitions from Schwarzschild to a Hartle--Thorne geometry \cite{Hartle1967,HartleThorne1968}, where light deflection acquires multipolar corrections \cite{EpsteinShapiro1980,IyerHansen2009}:
\begin{equation}
\hat{\alpha}(b) \approx \frac{4GM_{\rm tot}}{c^2 b} 
\mp \frac{4G J}{c^3 b^2} 
+ \mathcal{O}\!\left(\frac{G Q}{c^2 b^3}\right) ,
\end{equation}
where the upper and lower signs denote retrograde and prograde orbits, respectively, following the convention of Iyer and Hansen~\cite{IyerHansen2009}.

For the vast majority of isolated neutron stars ($P \gtrsim 100\text{ ms}$), the dimensionless spin parameter $a/M \equiv c J / (G M_{\rm tot}^2) \ll 0.1$ \cite{Urbanec2013}, rendering both frame-dragging ($\propto b^{-2}$) and quadrupole distortion ($\propto b^{-3}$) subdominant compared to the halo-induced deficit $\delta(b)$. Chiefly, inside the halo ($b < \Rd$), the metric is already non-Schwarzschild owing to the distributed dark matter mass profile $m(r)$; the extended physical volume of the halo ($\Delta r \sim 10\text{--}30\text{ km}$) means its flat-profile deflection signature dominates the rapidly decaying higher-multipole terms at intermediate impact parameters ($\Rb < b \le \Rd$).

For rapidly spinning millisecond pulsars ($P \lesssim 2\text{ ms}$), where $a/M \sim 0.1\text{--}0.3$, frame dragging introduces a measurable asymmetry between prograde and retrograde lensing trajectories. Extending the two-fluid optical metric to an axisymmetric Hartle--Thorne setup will allow future work to disentangle rotational multipoles from dark-sector mass profiles.

\begin{acknowledgments}
Y.~K. acknowledges that this work was supported by the Universidad Nacional Autónoma de México Postdoctoral Program (POSDOC). 
I.L. thanks the Funda\c{c}\~ao para a Ci\^encia e Tecnologia (FCT),
	Portugal, for financial support to the Centre for Astrophysics and
	Gravitation (CENTRA/IST/ULisboa) through grant No.~UID/PRR/00099/2025
	(\doi{10.54499/UID/PRR/00099/2025}) and grant No.~UID/00099/2025
	(\doi{10.54499/UID/00099/2025}).
\end{acknowledgments}$ $

\bibliographystyle{apsrev4-2}
\bibliography{ArtML}

\end{document}